\documentclass[twocolumn]{aastex63}

\usepackage{import}
\usepackage{xcolor}
\usepackage{comment}
\usepackage{ulem}

\graphicspath{ {./Images/} }

\shorttitle{A Study of Millimeter Variability in FUor Objects}
\shortauthors{Wendeborn et al.}

\begin{document}

\title{A Study of Millimeter Variability in FUor Objects}

\author{John Wendeborn}
\affil{Department of Astronomy \& Institute for Astrophysical Research, Boston University, 725 Commonwealth Avenue, Boston MA, 02215, USA}

\author{Catherine C. Espaillat}
\affil{Department of Astronomy \& Institute for Astrophysical Research, Boston University, 725 Commonwealth Avenue, Boston MA, 02215, USA}

\author{Enrique Mac\'ias}
\affil{Department of Astronomy \& Institute for Astrophysical Research, Boston University, 725 Commonwealth Avenue, Boston MA, 02215, USA}
\affil{Joint ALMA Observatory, Alonso de C\'ordova 3107, Vitacura, Santiago 763-0355, Chile}
\affil{European Southern Observatory, Alonso de C\'ordova 3107, Vitacura, Santiago 763-0355, Chile}

\author{Orsolya Feh\'er}
\affil{IRAM, 300 Rue de la Piscine, 38046 Saint Martin d'H\'eres, France}

\author{\'A. K\'osp\'al}
\affiliation{Konkoly Observatory, Research Centre for Astronomy and Earth Sciences, Konkoly-Thege Mikl\'os \'ut 15-17, 1121 Budapest, Hungary}
\affiliation{Max Planck Institute for Astronomy, K\"onigstuhl 17, 69117 Heidelberg, Germany}
\affiliation{ELTE E\"otv\"os Lor\'and University, Institute of Physics, P\'azm\'any P\'eter s\'et\'any 1/A, 1117 Budapest, Hungary}

\author{Lee Hartmann}
\affil{Department of Astronomy, University of Michigan, 500 Church Street, Ann Arbor MI, 41809, USA}

\author{Zhaohuan Zhu}
\affil{Department of Physics and Astronomy, University of Nevada, Las Vegas, 4505 South Maryland Parkway, Las Vegas NV, 89154, USA}

\author{Michael M. Dunham}
\affil{Department of Physics, State University of New York at Fredonia, 280 Central Avenue, Fredonia NY, 14063, USA}

\author{Marina Kounkel}
\affil{Department of Physics and Astronomy, Western Washington University, 516 High Street, Bellingham, WA 98225, USA}

\begin{abstract}

FU Orionis objects (FUors) are rapidly-accreting, pre-main sequence objects that are known to exhibit large outbursts at optical and near-infrared wavelengths, with post-eruption, small-scale photometric variability superimposed on longer-term trends. In contrast, little is known about the variability of FUors at longer wavelengths. To explore this further, we observed six FUor objects using the NOrthern Extended Millimeter Array (NOEMA) and for a subset of three objects we obtained coordinated observations with NOEMA and the Lowell Discovery Telescope (LDT). In combination with previously published NOEMA observations from 2014, our 2017 observations of V1735 Cyg provide the first detection of variability in an FUor object at 2.7 mm. In the absence of significant optical variability, we discount the possibility that the mm flux density changed as a result of irradiation from the central disk. In addition, a change in the dust mass due to infall is highly unlikely. A plausible explanation for the change in 2.7 mm flux density is variability in free-free emission due to changes in the object's jet/wind. Thus, it may be that free-free emission in some FUor objects is significant at $\sim$3 mm and must be considered when deriving disk masses in order to help constrain the mechanism responsible for triggering FUor outbursts.

\end{abstract}

\keywords{FUor objects --- star formation --- accretion --- protoplanetary disks --- variability}

\section{Introduction} 

FU Ori objects were originally identified as systems in star-forming regions that exhibit large outbursts at optical wavelengths \citep{Ambart1971, Herbig1977, Hartmann1996, Audard2014}. These outbursts have been explained by the onset of rapid mass accretion, rising from quiescent states of 10$^{-8}$--10$^{-7}$ M$_{\odot}$ yr$^{-1}$ to as much as 10$^{-4}$ M$_{\odot}$ yr$^{-1}$ \citep{Hartmann1996}. In their high states, FUors can also exhibit low-amplitude photometric/spectroscopic variability on $\sim$weekly timescales \citep{Siwak2018, Takagi2018}. Such high accretion rates, which can last for decades, may explain how stars accrete much of their mass, up to 10$^{-2}$ M$_{\odot}$ for a single outburst \citep{Hartmann1996}.

The causes of these outbursts are not clearly understood. Several explanations, including perturbations due to nearby companions, thermal instabilities, and gravitational instabilities, have all been proposed \citep{B&B1992, Clarke2005, PP6, Hartmann1996}. Companion perturbations, while an attractive solution for particular systems with specific orbital parameters, fail to explain many FUor events for isolated stars. Thermal and/or gravitational instabilities are more attractive alternatives for the underlying mechanism(s) behind FUor outbursts.

Thermal instabilities can occur when high disk opacities trap heat, leading to a runaway temperature increase within the disk. This heightened temperature then increases the effective disk viscosity, which in turn leads to high accretion rates \citep{Bell1994}. This is considered necessary to explain the fast rise times of FU Ori and V1057 Cyg during their initial outbursts \citep{Audard2014}. However, thermal instabilities depend significantly
on the disk viscosity and temperature, which are generally difficult to determine \citep{Vorobyov2005}.

Gravitational instabilities are thought to occur when mass infalling from the surrounding envelope  builds up in the disk, becoming gravitationally unstable. These instabilities can manifest in several ways, including magnetorotational instabilities \citep[e.g.][]{Armitage2001, Zhu2010} and/or clumps of material that then accrete onto the host star \citep[e.g.][]{Vorobyov2005}. If the envelope continues to dump mass into the disk over time, outbursts can be repetitive, something thought to be generally true of FUor objects \citep{Hartmann1985}.

Despite the above caveats, it is possible that both thermal and gravitational instabilities work in concert to produce FUor events. However, both explanations require disk masses large enough to sustain high mass accretion rates (10$^{-4}$ M$_\odot$ yr$^{-1}$) for decades, even 100's of years, and for the gravitational instability to be triggered \citep[M$_{disk}$/M$_*\gtrsim0.1$, see][]{Hartmann1996, Liu2016, Cieza2018}. Some measurements of FUor disk masses have called into question whether disks are, in general, massive enough for these instabilities to fully explain the observed outburst. For example, \citet{Dunham2012} and \citet{Kospal2016} found upper limits on the disk mass of the FUor object HBC 722 to be 0.02 M$_{\odot}$ and 0.01 M$_\odot$, respectively, likely too small for gravitational instabilities to explain its outburst \citep{Audard2014}. \citet[][hereafter F17]{Feher2017} also found that the disk masses of several FUor objects are quite low (e.g., 0.04 M$_{\odot}$ and $<$0.05 M$_{\odot}$ for V1515 Cyg and V710 Cas, respectively).

In order to gauge the viability of gravitational and/or thermal instabilities to explain FUor outbursts, accurate estimates of disk masses need to be made. This is not simple given that recent observations of FUor objects have shown that there may be up to 25--60\% variability at 1.3 mm \citep{Liu2018}. \citet{Liu2018} pointed to two possible explanations for the variability, including the perturbations of the thermal or density structure in the disk, as well as increased irradiation from the host star.

Coordinated millimeter and optical/NIR observations can provide insight into the underlying mechanism behind the observed millimeter variability. Most of the millimeter emission from FUor objects is thought to originate from the outermost regions of the disk, and traces cool, optically thin, millimeter-sized dust grains. The shorter optical/NIR emission originates from the innermost, hottest regions of the disk. A simultaneous change at millimeter and optical/NIR wavelengths would indicate that such variation is likely due to temperature changes in the disk, and hence thermal instabilities. This would further imply that future millimeter observations of FUor objects could, in general, benefit from coordinated optical/IR observations so as to better constrain the disk's properties \citep{PP6}. On the other hand, observing solely a change in millimeter flux density would indicate that the millimeter emission is disconnected from perturbations in the thermal structure of the disk, perhaps pointing towards gravitational instabilities or other alternative mechanisms.

In the following, we present millimeter observations of six FUor objects and coordinated millimeter and optical observations for a subset of three of these objects in order to determine the manner in which they vary over time. Details of our observations, as well as our data reduction techniques can be found in Section \ref{sec:obs}. Section \ref{sec: results&analysis} describes our analysis of these observations and the results. We discuss our findings in Section \ref{sec:disc}.

\section{Observations and Data Reduction} \label{sec:obs}

\subsection{Sample}

Our entire sample consists of six known FUor objects (V1735 Cyg, V2494 Cyg, V2495 Cyg, V1057 Cyg, V1515 Cyg, and V733 Cep). All six targets had one observing run at 2.7mm in May-June 2017. In this paper, we focus primarily on V1735 Cyg, V2494 Cyg, and V2495 Cyg, which were selected for two follow-up observing runs including both millimeter and optical observations in June and August 2018. V2494 Cyg and V2495 Cyg were chosen for follow-up observations because \citet{Liu2018} found tentative evidence for millimeter variability in these objects at 1.3 mm. V1735 Cyg was chosen because our 2017 data displayed variability relative to 2014 observations taken by F17. Our 2017 observations of V1057 Cyg, V1515 Cyg, and V733 Cep were consistent with 2014 flux densities reported by F17, so we elected not to carry out follow-up observations of these targets.

\subsection{Millimeter Observations} \label{sec: mm obs}

\begin{deluxetable*}{c c c c c c}[htp]
\setlength{\tabcolsep}{10pt}
\tablecaption{NOEMA Observation Log \label{tab:NOEMAObs}}
\centering
\tablehead{
\colhead{Object} & \colhead{RA/Dec} & \colhead{Date} & \colhead{Array Config.} & \colhead{Central Frequency} & \colhead{Distance$^a$} \\
\colhead{} & \colhead{(J2000)} & \colhead{} & \colhead{(No. of Antennas)} & \colhead{(GHz)} & \colhead{(pc)}
}
\startdata
V1735 Cyg & 21:47:20.66$^a$ & April 5, 2014  & 6Cq (6) & 109.918 & 616$^c$ \\
 & +47:32:03.6 & June 11/12/14, 2017 & 8D-E10 (7)/8D (8)/8D-W12N13 (6) & 108.502 & \\
 & & June 9/10, 2018 & 8ant-Special (8) & 106.744 & \\
 & & August 14/15, 2018 & 8D-W05 (7) & 106.744 & \\
\hline
V2494 Cyg & 20:58:21.09$^a$ & May 31/June 11, 2017 & 8D-N09 (7)/8D-E10 (7) & 108.502 & 800$^c$ \\
 & +52:29:27.7 & June 10, 2018 & 8ant-Special (8) & 106.744 & \\
 & & August 14/15, 2018 & 8D-W05 (7) & 106.744 & \\
\hline
V2495 Cyg & 21:00:25.38$^b$ & June 10, 2017 & 8D (8) & 108.502 & 800$^d$ \\
 & +52:30:15.5 & June 10, 2018 & 8ant-Special (8) & 106.744 & \\
 & & August 14/15, 2018 & 8D-W05 (7) & 106.744 & \\
\hline
V1057 Cyg & 20:58:53.73$^a$ & May 31/June 2, 2017 & 8D-N09 (7) & 108.502 & 898$^c$ \\
 & +44:15:28.38 & & & & \\
\hline
V1515 Cyg & 20:23:48.02$^a$ & June 3/4/7, 2017 & 8D-N09 (7) & 108.502 & 981$^c$ \\
 & +42:12:25.78 & & & & \\
\hline
V733 Cep & 22:53:33.26$^a$ & June 5/8, 2017 & 8D-N09 (7)/8D (8) & 108.502 & 661$^c$ \\
 & +62:32:23.63 & & & &
\enddata
\tablenotetext{}{$^a$\citet{Gaia}, $^b$\citet{Cutri2012}, $^c$\citet{Bailer2018}, $^d$\citet{Magakian2013}}
\end{deluxetable*}

Our millimeter observations were carried out using the NOrthern Extended Millimeter Array (NOEMA) in Plateau de Bure, France in May and/or June 2017 for all six objects in our sample. Three objects in our sample (V1735 Cyg, V2494 Cyg, and V2495 Cyg) were observed twice more in June 2018 and August 2018. This gave us baselines of one year and two months. We also used the April 2014 observations from F17 of V1735 Cyg, which gives us a longer, three-year baseline for that object. Observing conditions were generally good throughout each observation, with stable precipitable water vapor typically below 10 mm. Conditions during our 2017 observations were worse, with precipitable water vapor levels sometimes above 10 mm. More details of all observations can be found in Table \ref{tab:NOEMAObs}.

All continuum observations reported here were centered near 108 GHz ($\sim$2.8 mm), while the 2014 observations of V1735 Cyg from F17 were centered near 110 GHz ($\sim$2.7 mm). The 2017 observations used  NOEMA's WideX correlator tuned from 106.743--110.261 GHz, which covered the $^{13}$CO(1--0) (110.201 GHz) and C$^{18}$O(1--0) (109.782 GHz) lines at 78.118 kHz resolution. Our 2018 observations used the new PolyFix correlator tuned from 87.399 to 95.117 GHz (lower sideband) and 102.886 to 110.603 GHz (upper sideband). The two frequency ranges allowed us to measure the continuum emission at two different wavelengths, 2.81 mm and 3.29 mm. These frequencies covered the lines mentioned above, as well as the $^{13}$CS(2--1) (92.494 GHz) and HCO$^+$(1--0) (89.189 GHz) lines, at a spectral resolution of 62.495 kHz.

\begin{deluxetable}{c c c}[h!]
\setlength{\tabcolsep}{15pt}
\centering
\tablecaption{Summary of Observed Molecular Lines \label{tab:MolInfo} }
\tablehead{
\colhead{Molecule} & \colhead{Line Frequency} & \colhead{Transition}
}
\startdata
$^{13}$CO & 110.201 GHz & J=1-0 \\
C$^{18}$O & 109.782 GHz & J=1-0 \\
$^{13}$CS & 92.494 GHz & J=2-1 \\
HCO$^+$   & 89.189 GHz & J=1-0
\enddata
\end{deluxetable}

Data reduction was done in GILDAS \citep{Pety2005, Gildas2013} using the NOEMA pipeline in the CLIC package. Minimal flagging was required to remove spurious data. The calibrated data were then transferred to CASA 5.3.0 \citep{McMullin2007} for cleaning and further analysis. The flux calibrator MWC349 was used for all observations. J2037+511 was the phase calibrator for V2494 Cyg and V2495 Cyg, while J2201+508 was used for V1735 Cyg. The flux calibration source MWC 349 is time variable at 3 mm by $<$10\%. Taking this and other factors into account, the nominal absolute flux uncertainty of NOEMA at 2.7 mm is estimated to be 10\%.

We note that we cannot exclude the possibility that unaccounted instrumentation errors also affected the flux calibration (e.g., antenna pointing errors). We expect any other possible effect to have a lower impact than the nominal 10\% absolute flux calibration uncertainty. However, quantifying their impact is extremely uncertain, and could lead to an underestimate of all of our cited flux calibration uncertainties.

Continuum data of V1735 Cyg, V2494 Cyg, V2495 Cyg were first restricted to a \textit{uv} range of 30--65 k$\lambda$ in order to mitigate the effects of different uv-coverages and ensure that the amount of missing flux is the same between epochs. This also has the effect of removing extended emission, thereby ensuring that the measured flux densities are solely of the central, compact source. These continuum data were imaged using the \textit{clean} task with natural weighting and then convolved to a $4\arcsec\times4\arcsec$ beam. After cleaning, but before uv-restriction/beam convolution, our angular resolution was about 3-4$\arcsec$. After restricting the \textit{uv} range to 30--65 k$\lambda$, our resolutions improved to about 2-3$\arcsec$. And after beam convolution, the resolution was constant at 4$\arcsec$.

The resulting continuum images show compact emission for all the sources. The morphology of this continuum emission remained unchanged. No source was resolved, either fully or marginally, at 2.7 mm, before or after \textit{uv} restriction or beam convolution. Continuum data of V1057 Cyg, V1515 Cyg, and V733 Cep were cleaned in the same manner, but were not uv-restricted nor convolved to a $4\arcsec\times4\arcsec$ beam. The line data were also processed using the \textit{clean} task and natural weighting, but were corrected using \textit{imcontsub} to remove continuum emission. The spectral resolution for all data cubes was about 0.25 km/s. Continuum rms values were obtained from emission-free regions, whereas line rms values were determined using emission-free channels. Note that the continuum rms values therefore include the effects of thermal and phase atmospheric noise as well as some contribution from the imaging reconstruction process due to the limited uv-coverage, whereas the line rms values do not include the latter effect.

\begin{deluxetable*}{ccccc}[h!]
\setlength{\tabcolsep}{15pt}
\tablecaption{Millimeter Continuum Flux Densities \label{tab:ContFluxes} }
\centering
\tablehead{
\colhead{Object} & \colhead{Date} & \colhead{I$_{\nu,Peak}$} & \colhead{S$_{\nu,Gauss}$} & \colhead{rms} \\
\colhead{} & \colhead{} & \colhead{(mJy/beam)} & \colhead{(mJy)} & \colhead{(mJy/beam)}
}
\startdata
V1735 Cyg & April 2014 & 1.63 $\pm$ 0.16$^a$ & 1.33 $\pm$ 0.16$^a$ & 0.10 \\
& June 2017 & 2.27 $\pm$ 0.34$^a$ & 2.44 $\pm$ 0.39$^a$ & 0.07 \\
& June 2018 & 1.94 $\pm$ 0..19 & 1.84 $\pm$ 0.20 & 0.08 \\
& August 2018 & 1.68 $\pm$ 0.17 & 1.60 $\pm$ 0.19 & 0.09 \\
\hline
V2494 Cyg & May/June 2017 & 16.7 $\pm$ 2.5$^a$  & 16.6 $\pm$ 2.5$^a$ & 0.16 \\
& June 2018 & 16.0 $\pm$ 1.6 & 15.8 $\pm$ 1.6 & 0.07 \\
& August 2018 & 13.8 $\pm$ 1.4 & 12.5 $\pm$ 1.4 & 0.40 \\
\hline
V2495 Cyg & June 2017 & 14.6 $\pm$ 2.2$^a$ & 13.9 $\pm$ 2.2$^a$ & 0.17 \\
& June 2018 & 14.0 $\pm$ 1.4 & 13.8 $\pm$ 1.4 & 0.07 \\
& August 2018 & 12.9 $\pm$ 1.3 & 12.2 $\pm$ 1.2 & 0.24 \\
\hline
V1057 Cyg & May/June 2017 & 3.97 $\pm$ 0.60 & 5.40 $\pm$ 0.81 & 0.08 \\
\hline
V1515 Cyg & June 2017 & 0.66 $\pm$ 0.10 & 0.84 $\pm$ 0.13 & 0.04 \\
\hline
V733 Cep & June 2017 & 0.54 $\pm$ 0.08 & 1.50 $\pm$ 0.23 & 0.05 
\enddata
\tablenotetext{a}{Corrected for frequency discrepancy, using a spectral index of 2.5}
\tablecomments{ I$_{\nu,Peak}$ and S$_{\nu,Gauss}$ are obtained from Gaussian fitting and the rms is obtained from an emission-free region.}
\end{deluxetable*}

\begin{deluxetable*}{cccccccc}[htp]
\setlength{\tabcolsep}{7.5pt}
\tablecaption{$^{13}$CO, C$^{18}$O Fluxes \label{tab:13COC18OFluxes} }
\centering
\tablehead{ \colhead{Object} & \colhead{Date} & \colhead{I$_{\nu,13CO}$} & \colhead{rms$_{13CO}$} & \colhead{$\Delta$v$_{13CO}$} & \colhead{I$_{\nu,C18O}$} & \colhead{rms$_{C18O}$} & \colhead{$\Delta$v$_{C18O}$} \\ 
\colhead{} & \colhead{} & \colhead{(Jy km/s)} & \colhead{(Jy/beam km/s)} & \colhead{(km/s)} & \colhead{(Jy km/s)} & \colhead{(Jy/beam km/s)} & \colhead{(km/s)}
}
\startdata
V1735 Cyg & April 2014 & 0.7 $\pm$ 0.2 & 0.12 & --4.43, +5.87 & 0.21 $\pm$ 0.05 & 0.03 & --2.73, +1.52 \\
& June 2017 & 0.7 $\pm$ 0.2 & 0.12 & --4.50, +6.00 & 0.22 $\pm$ 0.06 & 0.03 & --2.75, +1.50 \\
& June 2018 & 1.0 $\pm$ 0.3 & 0.11 & --4.50, +6.00 & 0.21 $\pm$ 0.05 & 0.04 & --2.75, +1.50 \\
& August 2018 & 1.3 $\pm$ 0.3 & 0.17 & --4.50, +6.00 & 0.39 $\pm$ 0.10 & 0.06 & --2.75, +1.50 \\
\hline 
V2494 Cyg & May/June 2017 & 1.2 $\pm$ 0.3 & 0.09 & --2.75, +2.25 & 0.40 $\pm$ 0.10 & 0.05 & --3.00, +1.50 \\
& June 2018 & 1.2 $\pm$ 0.3 & 0.08 & --2.75, +2.25 & 0.40 $\pm$ 0.10 & 0.04 & --3.00, +1.50 \\
& August 2018 & 1.2 $\pm$ 0.3 & 0.11 & --2.75, +2.25 & 0.38 $\pm$ 0.10 & 0.06 & --3.00, +1.50 \\
\hline
V2495 Cyg & June 2017 & 0.19 $\pm$ 0.05 & 0.03 & --2.00, +3.75 & 0.22 $\pm$ 0.06 & 0.03 & --2.00, +2.50 \\
& June 2018 & 0.23 $\pm$ 0.06 & 0.05 & --2.00, +3.75 & 0.28 $\pm$ 0.07 & 0.04 & --2.00, +2.50 \\
& August 2018 & 0.31 $\pm$ 0.08 & 0.03 & --2.00, +3.75 & 0.27 $\pm$ 0.07 & 0.05 & --2.00, +2.50 \\
\hline
V1057 Cyg & May/June 2017 & 2.2 $\pm$ 0.6 & 0.15 & --4.32, +1.84 & 0.49 $\pm$ 0.12 & 0.03 & --2.12, +1.08 \\
\hline 
V1515 Cyg$^a$ & June 2017 & -- & -- & -- & -- & -- & -- \\
\hline 
V733 Cep$^a$  & June 2017 & -- & -- & -- & -- & -- & --
\enddata
\tablenotetext{a}{ No data due to poor quality}
\tablecomments{ I$_{\nu}$ is the flux obtained from a $5.66\arcsec\times5.66\arcsec$ aperture on the object location. rms$_{13CO}$ and rms$_{C18O}$ are the background rms obtained from emission-free regions. $\Delta$v is the integrated velocity range.}
\end{deluxetable*}

\begin{deluxetable*}{c c c c c c c c}[htp]
\setlength{\tabcolsep}{7.5pt}
\tablecaption{$^{13}$CS, HCO$^+$ Fluxes \label{tab:13CSHCO+Fluxes} }
\centering
\tablehead{ \colhead{Object} & \colhead{Date} & \colhead{I$_{\nu,13CS}$} & \colhead{rms$_{13CS}$} & \colhead{$\Delta$v$_{13CS}$} & \colhead{I$_{\nu,HCO^+}$} & \colhead{rms$_{HCO^+}$} & \colhead{$\Delta$v$_{HCO^+}$} \\ 
\colhead{} & \colhead{} & \colhead{(Jy km/s)} & \colhead{(Jy/beam km/s)} & \colhead{(km/s)} & \colhead{(Jy km/s)} & \colhead{(Jy/beam km/s)} & \colhead{(km/s)}
}
\startdata
V1735 Cyg & June 2018 & 0.033 $\pm$ 0.009 & 0.008 & --1.00, +0.50 & 1.14 $\pm$ 0.29 & 0.07 & --1.00, +7.50 \\
& August 2018 & 0.040 $\pm$ 0.010 & 0.014 & --1.00, +0.50 & 1.35 $\pm$ 0.34 & 0.11 & --1.00, +7.50 \\
\hline
V2494 Cyg & May/June 2018 & -- & 0.004 & --1.50, --0.50 & 0.18 $\pm$ 0.05 & 0.01 & --2.25, +0.25 \\
& August 2018 & -- & 0.007 & --1.50, --0.50 & 0.17 $\pm$ 0.04 & 0.03 & --2.25, +0.25 \\
\hline
V2495 Cyg & June 2018 & -- & 0.01 & --1.25, +3.25 & -- & 0.02 & --2.00, +1.00 \\
& August 2018 & -- & 0.03 & --1.25, +3.25 & -- & 0.03 & --2.00, +1.00 
\enddata
\tablecomments{ Our 2017 correlator setup did not include these lines. Therefore, the 2017 observations are not included in this table. I$_{\nu}$ is the flux obtained from a $5.66\arcsec\times5.66\arcsec$ aperture on the object location. rms$_{13CS}$ and rms$_{HCO^+}$ are the background rms obtained from emission-free regions. $\Delta$v is the integrated velocity range. We do not detect $^{13}$CS emission from V2494 Cyg or V2495 Cyg. We do not detect HCO$^+$ emission from V2495 Cyg.}
\end{deluxetable*}

\subsection{Optical Observations} \label{sec:opt obs}

\begin{deluxetable*}{c c c c c c c c c}[htp]
\setlength{\tabcolsep}{15pt}
\tablecaption{LDT Photometry \label{tab:OptObsMags} }
\centering
\tablehead{
\colhead{Object} & \colhead{Date} & \colhead{Start Time (UT)} & \colhead{V} & \colhead{R} & \colhead{I} }
\startdata
V1735 Cyg & June 11, 2018 & 10:44 & 19.02 $\pm$ 0.09 & 16.61 $\pm$ 0.06 & 14.24 $\pm$ 0.04 \\
& June 12, 2018 & 10:45 & 19.07 $\pm$ 0.09 & 16.61 $\pm$ 0.06 & 14.24 $\pm$ 0.04 \\
& August 13, 2018 & 11:10 & 19.13 $\pm$ 0.09 & 16.62 $\pm$ 0.07 & 14.28 $\pm$ 0.05 \\
\hline
V2494 Cyg & June 11, 2018 & 10:51 & 18.47 $\pm$ 0.09 & 16.65 $\pm$ 0.07 & 14.79 $\pm$ 0.04 \\
& June 12, 2018 & 10:53 & 18.47 $\pm$ 0.09 & 16.63 $\pm$ 0.06 & 14.80 $\pm$ 0.04 \\
& August 13, 2018 & 11:30 & 18.38 $\pm$ 0.09 & 16.55 $\pm$ 0.07 & 14.73 $\pm$ 0.05 \\
\enddata
\end{deluxetable*}

Optical observations of V1735 Cyg, V2494 Cyg, and V2495 Cyg were carried out using the Lowell Discovery Telescope's (LDT) Large Monolithic Imager (LMI) in Happy Jack, AZ \citep{Bida2014}, in June and August 2018. These observations were coordinated with our 2018 NOEMA observations. Our June millimeter and optical observations were separated by about 24 hours, whereas our August observations were separated by about 36 hours. Details of these observations can be found in Table \ref{tab:OptObsMags}.

All observations were performed in photometric conditions using three optical filters: Johnson V, R, and I, with central wavelengths of 551 nm, 658 nm, and 806 nm respectively \citep{Johnson1966}. The FOV of the LMI is 12.5$\arcmin$ $\times$ 12.5$\arcmin$ (0.12$\arcsec$ per unbinned pixel). We used 2 $\times$ 2 pixel binning for these observations (0.24$\arcsec$ per pixel). The bias, flat-field, and overscan calibration of the CCD images were performed using a custom Python routine utilizing the Astropy package \citep{Astropy2018}. The photometric calibration of all images was carried out interactively also using a custom Python routine utilizing the Astropy package. Exposure times varied among the targets, dependent on precise seeing conditions and object brightness. V2495 Cyg was too faint to detect in any of our observations. We were therefore unable to extract optical magnitudes for this object.

Using the standard stars SA92 253 and GD 246 \citep{Landolt2009}, we obtained V, R, and I magnitudes for our targets in August 2018. To obtain magnitudes for our June 2018 observations, we utilized differential photometry. This involved normalizing the flux density of our target to that of the total brightness of several background stars, which fluctuated by less than 3\% throughout the observations. Then we directly compared the June 2018 flux densities to those of August 2018. Finally, we used that ratio to obtain an optical magnitude for the June 2018 epoch. Magnitudes of our sample are listed in Table \ref{tab:OptObsMags}.

\section{Results and Analysis} \label{sec: results&analysis}

In the following, we present millimeter flux densities for all six FUors in our sample. We then search for variability in the 2.7 mm flux densities; $^{13}$CO, C$^{18}$O, $^{13}$CS, and HCO$^+$ molecular line fluxes; and optical magnitudes of V1735 Cyg, V2494 Cyg, V2495 Cyg. We then describe the variability that is seen in V1735 Cyg for the 2.7 mm continuum.

\subsection{Millimeter Continuum} \label{sec:mmresults}

Continuum flux densities were measured by 2D Gaussian fits within a $5.66\arcsec\times5.66\arcsec$ circular region, twice the convolved beam area. These results can be found in Table \ref{tab:ContFluxes}. To account for the frequency discrepancy, we adjusted our 2014 and 2017 measured flux densities using a spectral index of 2.5. The uncertainties stated in Table \ref{tab:ContFluxes} were obtained via the root sum square of the 10\% absolute flux calibration uncertainty of NOEMA (15\% in the case of June 2017 observations) and the Gaussian fit uncertainties (typically 1--5\%).

\begin{figure*}[t]
    \centering
    \includegraphics[width=\textwidth]{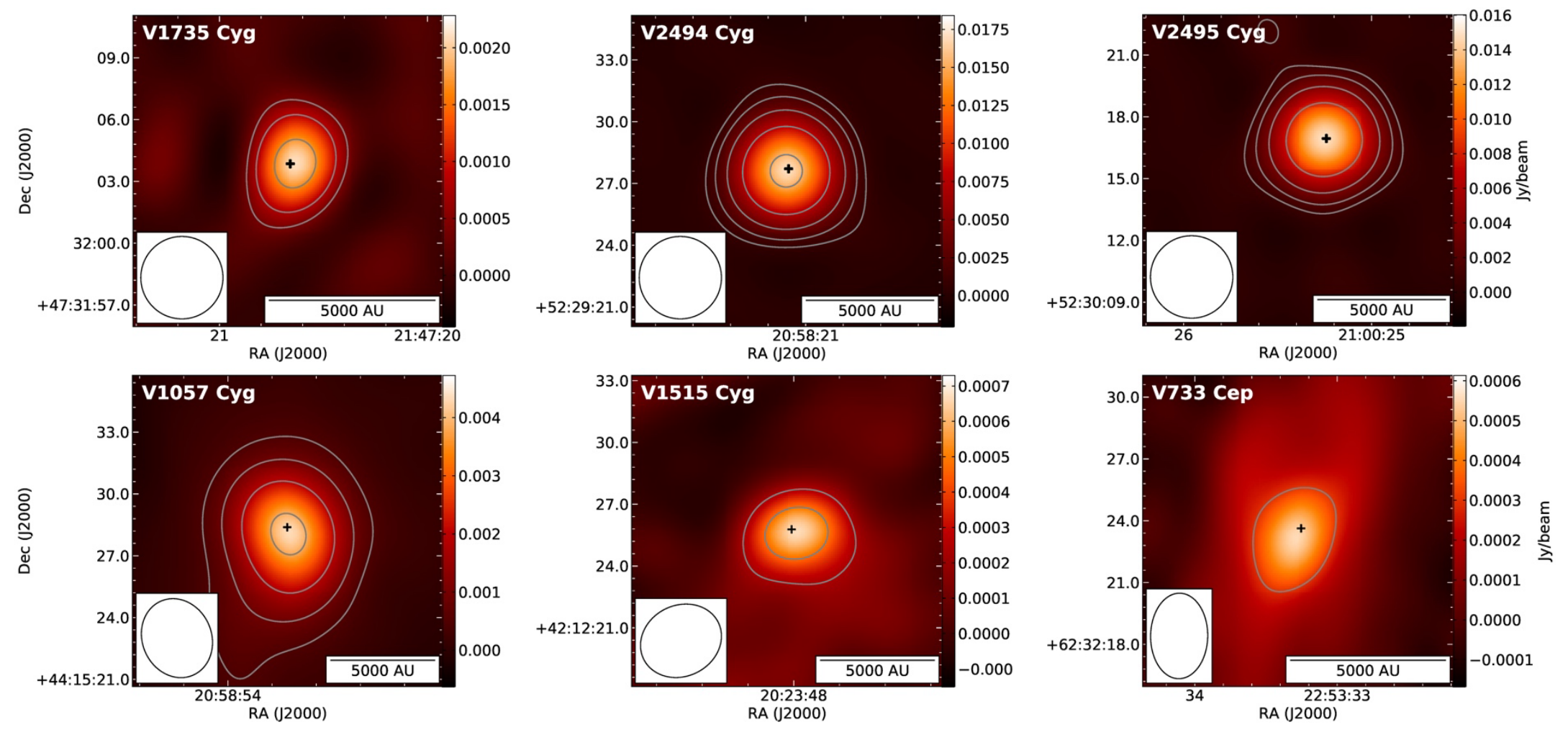}
    \caption{Continuum maps of all six sources. Solid contours denote positive 6-, 12-, 24-, 48-, 96-, 192-$\sigma$ levels. $\sigma$ levels are rms values noted in Table \ref{tab:ContFluxes}. The central `+' denotes object position (see Table \ref{tab:NOEMAObs}).}
    \label{fig:ContFigs}
\end{figure*}

Figure \ref{fig:ContFigs} shows the continuum maps of all six targets. While flux densities did vary somewhat, intensity distributions remained unchanged, thus we only show one epoch. The maps of V1735 Cyg, V2494 Cyg, V2495 Cyg were made using a restricted \textit{uv} coverage, whereas the rest were not. We note that despite steps taken to mitigate the differing \textit{uv} coverages, a portion of the variability reported may still be due to remaining differences. 

Also of note, our measured flux density for V1057 Cyg ($5.4\pm0.1$ mJy) agrees with that of F17 ($4.9\pm0.2$ mJy). V1515 Cyg and V733 Cep were weakly detected in F17, where F17 estimates peak intensities of 0.18 $\pm$ 0.03 and 0.38 $\pm$ 0.10 mJy/beam, respectively. F17 could not, however, estimate flux densities. The F17 peak intensities are somewhat lower than what we report here for V1515 Cyg and V733 Cep, 0.66 $\pm$ 0.10 and 0.54 $\pm$0.08 mJy/beam, respectively.

\begin{figure}[ht]
\centering
\includegraphics[width=.45\textwidth]{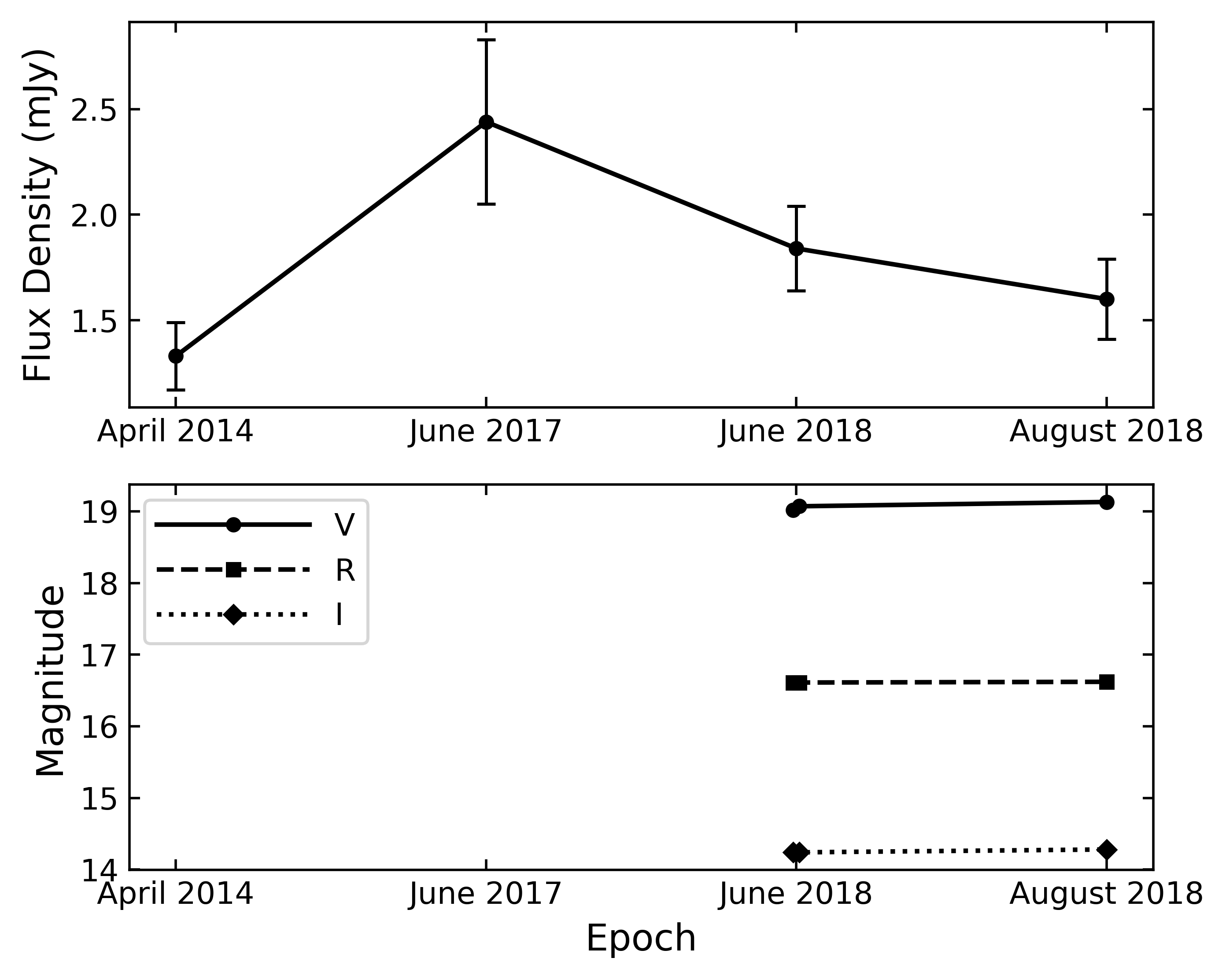}
\caption{Top: V1735 Cyg millimeter continuum flux density vs.\ time. Error bars are the uncertainties and are listed in Table \ref{tab:ContFluxes}. Bottom: V1735 Cyg VRI magnitudes vs.\ time. Error bars are roughly the size of the points.}
\label{fig:V1735mmVtime}
\end{figure}

\begin{figure}[ht]
\centering
\includegraphics[width=.45\textwidth]{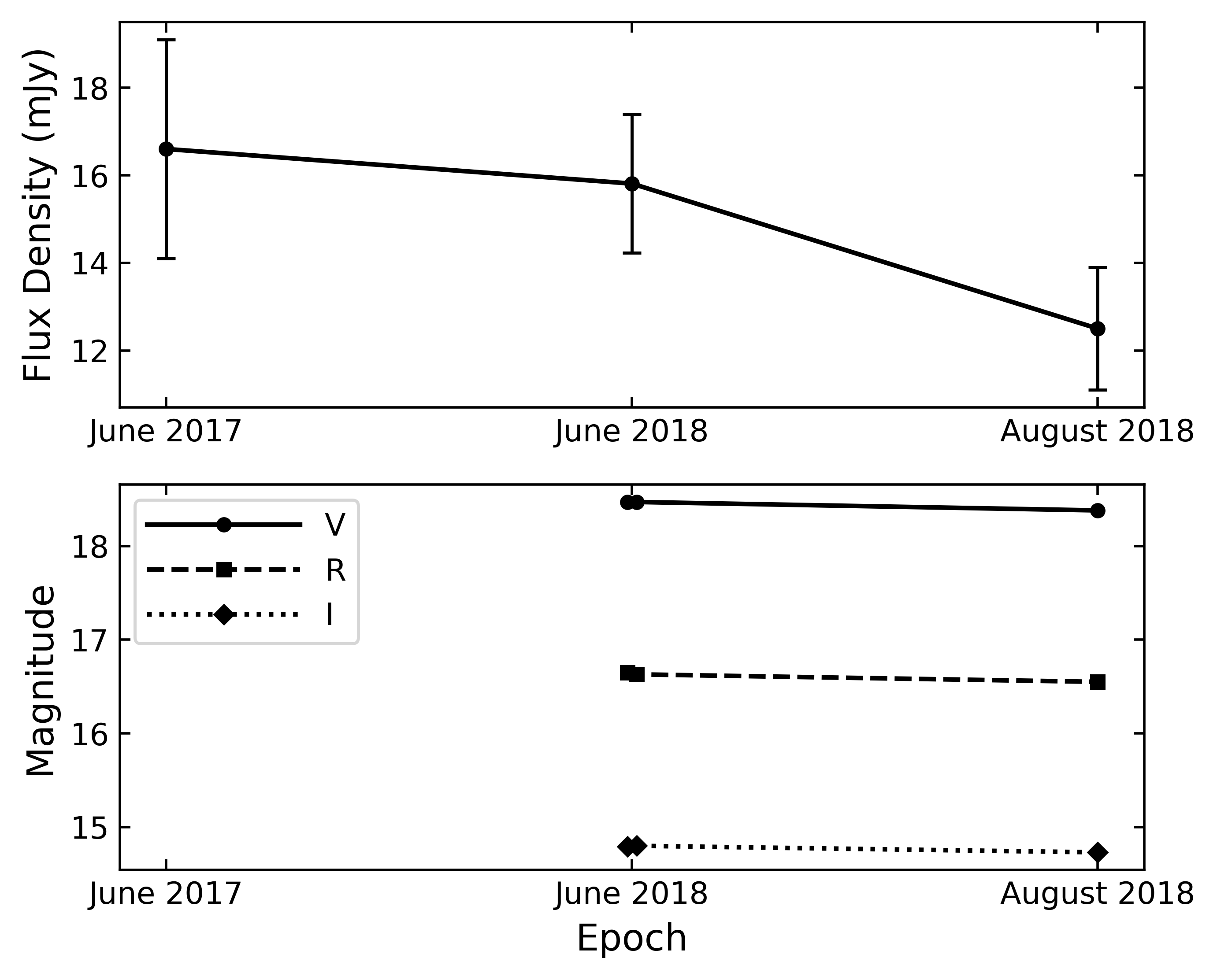}
\caption{Top: V2494 Cyg millimeter continuum flux density vs.\ time. Error bars are the uncertainties and are listed in Table \ref{tab:ContFluxes}. Bottom: V2494 Cyg VRI magnitudes vs.\ time. Error bars are roughly the size of the points.}
\label{fig:V2494mmVtime}
\end{figure}

\begin{figure}[ht]
\centering
\includegraphics[width=.45\textwidth]{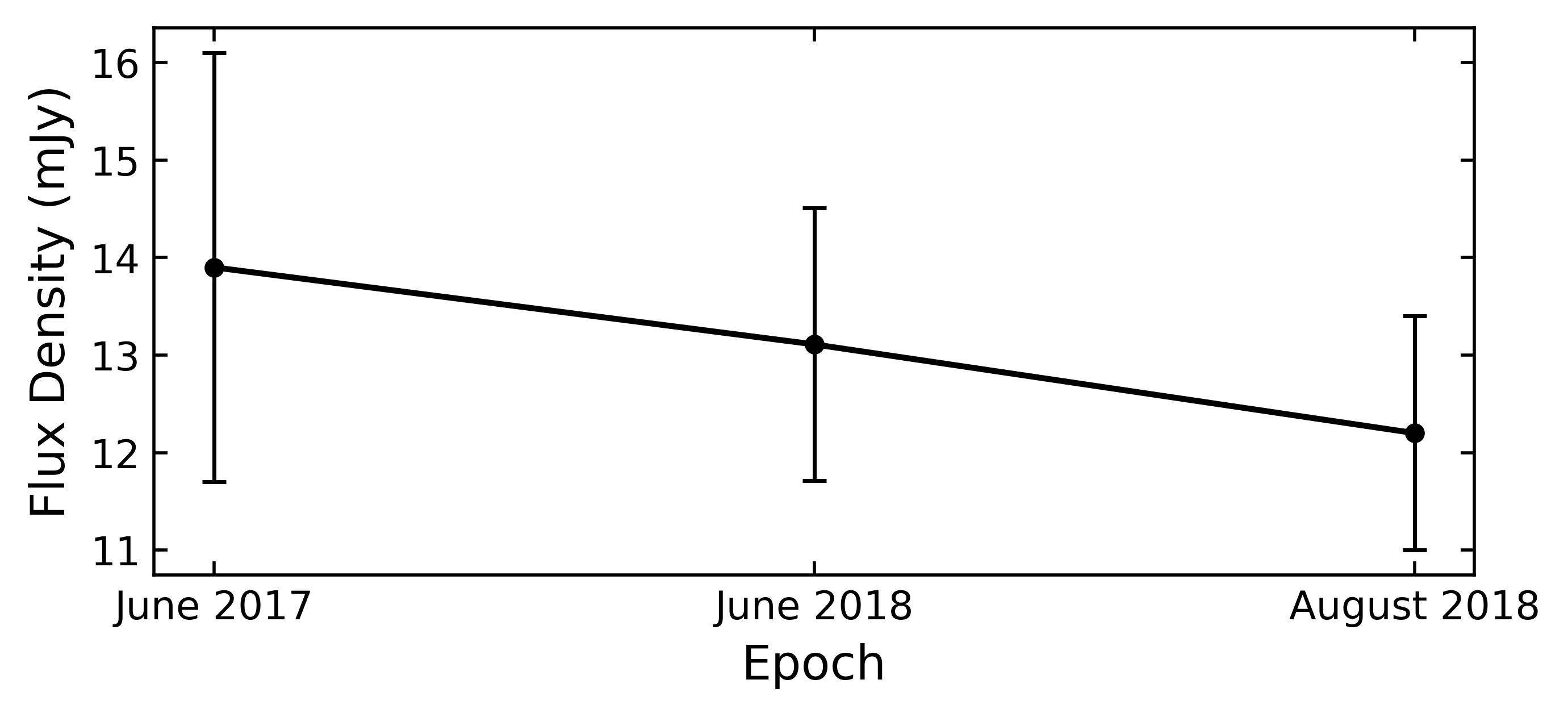}
\caption{V2495 Cyg millimeter continuum flux density vs.\ time. Error bars are the uncertainties and are listed in Table Table \ref{tab:ContFluxes}.} 
\label{fig:V2495mmVtime}
\end{figure}

For V1735 Cyg, V2494 Cyg, and V2495 Cyg, we show the continuum flux densities and optical magnitudes for the two epochs in 2018 in Figures \ref{fig:V1735mmVtime}, \ref{fig:V2494mmVtime}, and \ref{fig:V2495mmVtime}, respectively. The only object to display any millimeter variability in our sample, V1735 Cyg, exhibited an $\sim80$\% increase in flux density from 2014 to 2017. This flux density increase falls outside our stated uncertainties, thus, we conclude that the observed variability is intrinsic to the source.

We also see that following its rise from 2014 to 2017, V1735 Cyg possibly dimmed in June 2018 and then again in August 2018. This may be a sign that it is returning to some quiescent state, from some ``burstlike," heightened state. However, this downward trend from 2017 to 2018 is within or close to the flux uncertainties of NOEMA, and is also seen in both V2494 Cyg and V2495 Cyg. More observations are needed to confirm if V1735 Cyg's flux density at 3 mm has truly decreased since June 2017. No other objects in our sample, over any time period, show signs of millimeter variability outside of the measurement and flux calibration uncertainties. 

\subsection{Molecular Lines}

$^{13}$CO, C$^{18}$O, $^{13}$CS, and HCO$^+$ line fluxes for V1735 Cyg, V2494 Cyg, and V2495 Cyg were extracted from velocity-integrated spectral cubes after continuum subtraction and cleaning. The velocity range used for integration varied per object and per species, but was always centered in the frames that contained emission. In all cases, the same $5.66\arcsec\times5.66\arcsec$ circular aperture (centered at the primary source of emission) was used to measure the flux. Given the extended and asymmetric morphology of the line emission, we chose not to use 2D Gaussian fits to obtain line fluxes. These results can be found in Tables \ref{tab:13COC18OFluxes} and \ref{tab:13CSHCO+Fluxes}. We note that for the line fluxes, due to the variable \textit{uv} coverages, lower SNR, extension of the emission, and possible contamination from the surrounding envelope, we assume uncertainties of 25\%. In one case (V2495 Cyg) $^{13}$CS emission was not detected on or near the target's location, but further from the target at about $\sim15\arcsec$ away. We measure 0.039 and 0.051 Jy km/s in June and August 2018, respectively.

\begin{figure}[h!]
    \centering
    \includegraphics[width=.45\textwidth]{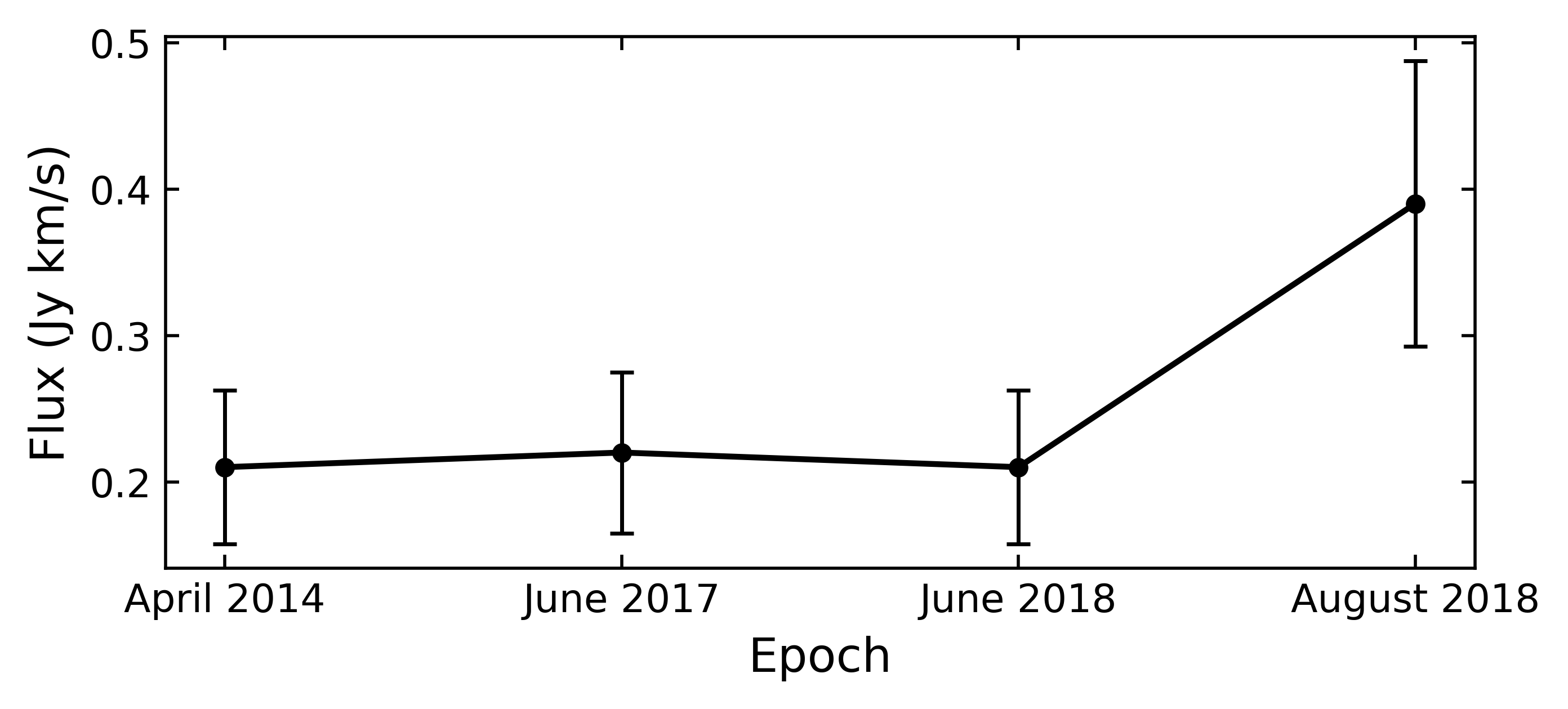}
    \caption{V1735 Cyg C$^{18}$O flux vs.\ time. Error bars are the uncertainties and are listed in Table \ref{tab:13COC18OFluxes}.}
    \label{V1735C18OVtime}
\end{figure}

$^{13}$CO and C$^{18}$O fluxes (Table \ref{tab:13COC18OFluxes}) were generally consistent for all objects across all epochs. The only exception may be the C$^{18}$O emission of V1735 Cyg from June 2018 to August 2018 (see Figure \ref{V1735C18OVtime}). The flux appears to have risen by about 86\%, though we emphasize that the line fluxes are highly uncertain because of the lack of short baselines to recover the extended emission. Additionally, the slightly different \textit{uv} coverages between epochs also result in artificial differences in the morphology of the extended line emission. Regardless, these differences have a relatively small (yet hard to quantify) effect on our flux measurements since we focus only on the compact line emission at the position of each object. This is partly shown by the fact that the 2014 observations (which included IRAM 30m observations to cover short \textit{uv} spacings) display similar fluxes to our observations.

\begin{figure*}[h!]
    \centering
    \includegraphics[width=\textwidth]{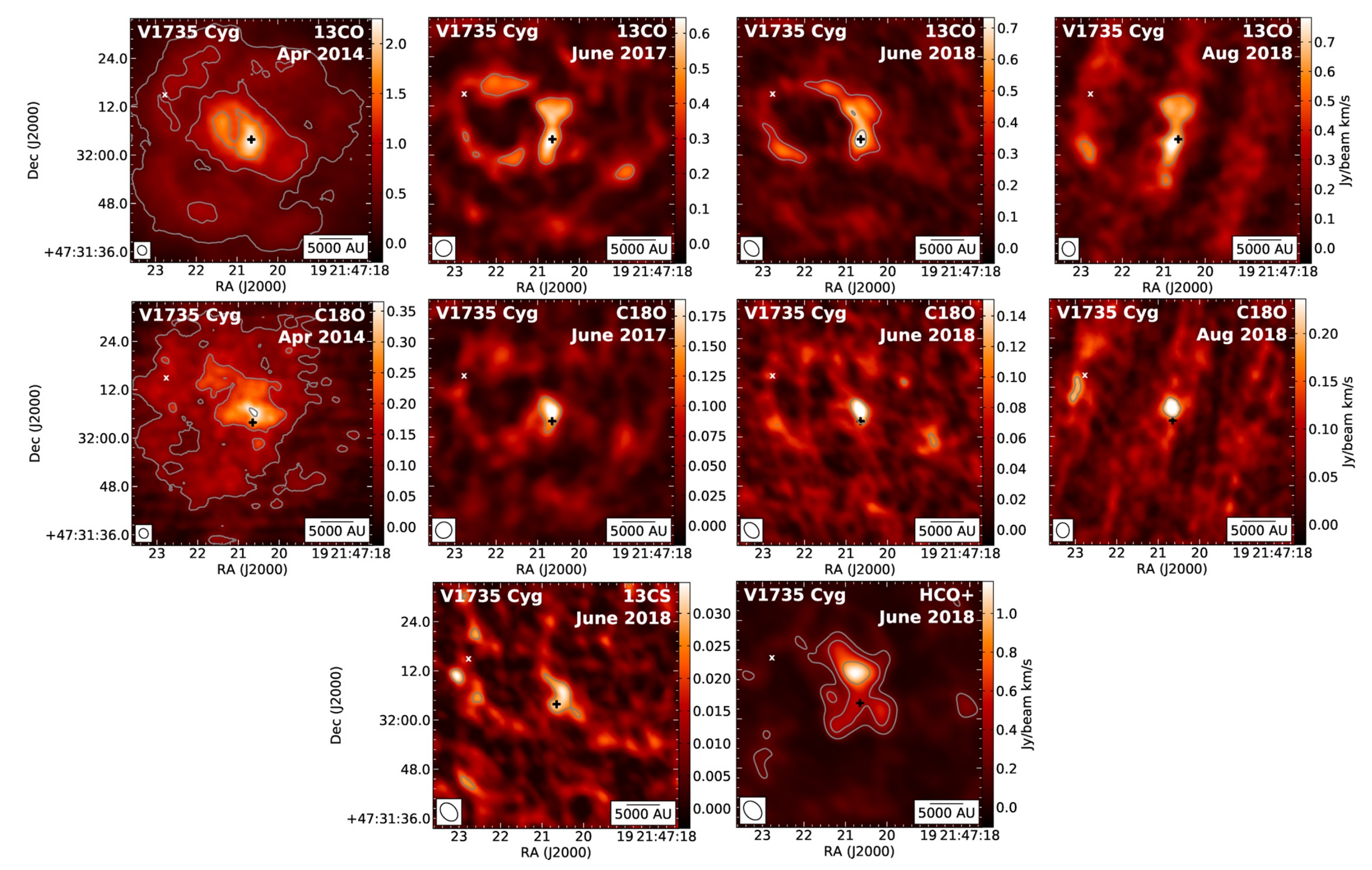}
    \caption{Moment-0 molecular line maps of V1735 Cyg. Top: $^{13}$CO. Middle: C$^{18}$O. Bottom: $^{13}$CS (left) and HCO$^+$ (right). Solid (dashed) gray contours denote positive (negative) 3-, 6-, 12-, 24-, 48-, 96-, and 192-$\sigma$ levels. $\sigma$ for each epoch is equivalent to the rms of each image, which can be found in Table \ref{tab:13COC18OFluxes}. The central `+' denotes the target's location (see Table \ref{tab:NOEMAObs}). The `$\times$' denotes the location of V1735 Cyg SM1 \citep{Harvey2008}. Morphological differences between April 2014 and other epochs is due to differing \textit{uv} coverages. Note that we do not include the $^{13}$CS and HCO$^+$ line maps from August 2018 since they are very similar to those of June 2018 shown here.}
    \label{fig:V1735LineMaps}
\end{figure*}

\begin{figure*}[h]
    \centering
    \includegraphics[width=\textwidth]{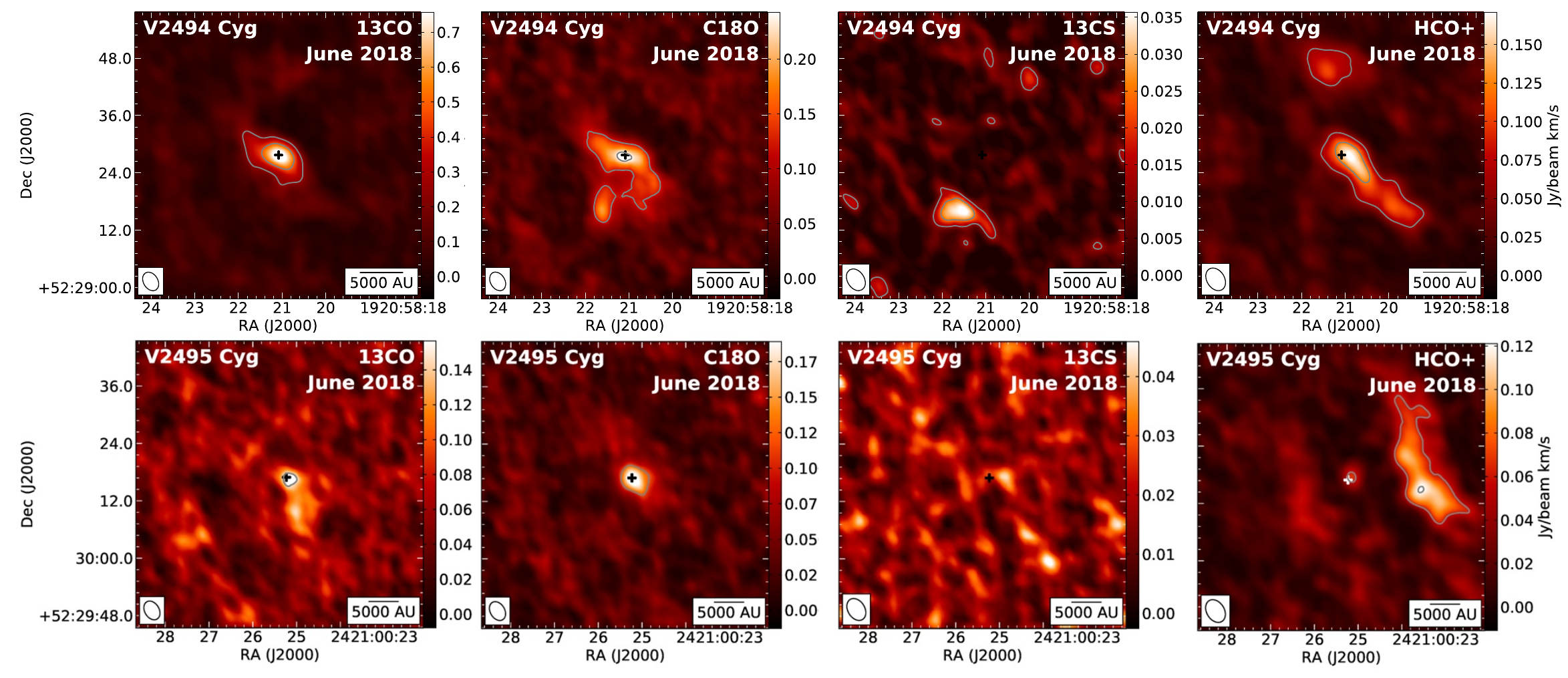}
    \caption{Moment-0 molecular line maps of V2494 Cyg (top) and V2495 Cyg (bottom). From left to right: $^{13}$CO, C$^{18}$O, $^{13}$CS, and HCO$^+$. Solid (dashed) gray contours denote positive (negative) 3-, 6-, 12-, 24-, 48-, 96-, and 192-$\sigma$ levels. $\sigma$ for each epoch is equivalent to the rms of each image, which can be found in Table \ref{tab:13COC18OFluxes}. The central `+' denotes the source's location (see Table \ref{tab:NOEMAObs}). Note that here we only show $^{13}$CO and C$^{18}$O line maps from June 2018 since the maps from June 2017 and August 2018 are very similar. Likewise, we only show line maps of $^{13}$CS and HCO$^+$ from June 2018 since the maps from August 2018 are very similar.}
    \label{fig:V2494V2495LineMaps}
\end{figure*}

$^{13}$CS and HCO$^+$ was observed in V1735 Cyg, V2494 Cyg, V2495 Cyg in June and August of 2018 (Table \ref{tab:13CSHCO+Fluxes}). These lines show no signs of variability, either in flux or spatial morphology. We did not detect $^{13}$CS or HCO$^+$ in V2495 Cyg. The emission morphology for $^{13}$CS and HCO$^+$ for all objects was consistent throughout 2018, thus we display only one epoch in Figures \ref{fig:V1735LineMaps} and \ref{fig:V2494V2495LineMaps}.

\subsection{LDT-LMI Photometry}

Optical photometry taken in June and August 2018 for V1735 Cyg and V2494 Cyg do not show variability (Table \ref{tab:OptObsMags}) and are consistent with previous observations. Our measurements of V1735 Cyg generally agree with those of \citet{Peneva2009} from 2003 to 2009. They measured V $\sim18.9$ and R $\sim16.6$, but found I $\sim13.8$, about half a magnitude brighter than reported here. Our measurements of V2494 Cyg agree with those of \citet{Magakian2013} from 2003 to 2010 in R and I. They found R $\sim16.4$ and I $\sim14.7$, but did not report V. We note that V2495 Cyg was observed, but not detected (see Section \ref{sec:opt obs}).

\section{Discussion} \label{sec:disc}

Prior to our observations, V2494 Cyg and V2495 Cyg were the only two FUor objects thought to be variable at millimeter wavelengths, displaying 1.3 mm flux density changes of $\sim$25--60\% on a timescale of about one year \citep{Liu2018}. Here we report that V1735 Cyg has also exhibited variability in the millimeter, but at 2.7 mm, and over a timescale of $\sim3$ years, from 2014 to 2017. We discuss here possible underlying mechanisms for this variability.

\subsection{Variable Disk Heating} \label{sec:heating}

In FUors, the inner disk is significantly heated by viscous heating from the accretion process and produces strong optical/IR emission \citep{Hartmann1996} and possibly millimeter emission as well \citep{Takami2019}. This hot inner disk irradiates the outer disk. Therefore, changes in the temperature of the inner disk may lead to changes in the heating of the outer disk, which we can trace with millimeter emission.

\begin{figure}[h!]
    \centering
    \includegraphics[width=.5\textwidth]{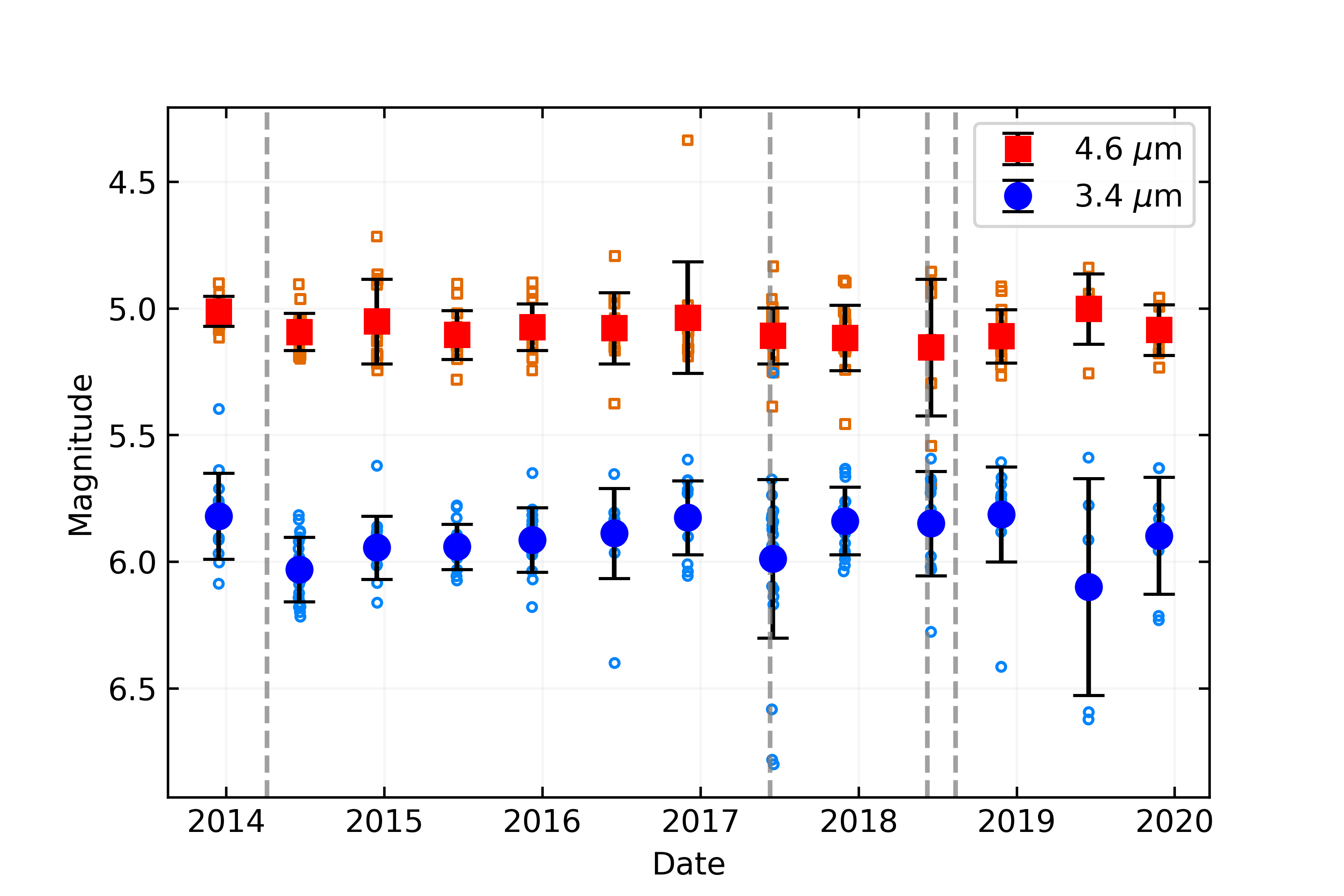}
    \caption{WISE photometry of V1735 Cyg from 2014--2019. Red circles are Band 1 (3.4 $\mu$m). Blue squares are Band 2 (4.6 $\mu$m). Dashed grey bars indicate dates of NOEMA observations.}
    \label{fig:WISE}
\end{figure}

If temperature changes in the inner disk were the cause of the millimeter variability we see in V1735 Cyg, we would also expect to see a corresponding increase in its optical and/or IR flux as well. To the best of our knowledge, there is no optical data of V1735 Cyg taken close in time to the 2014 NOEMA data and so we cannot test this using our 2018 optical data. However, archival WISE data of V1735 Cyg exists at 3.4 micron and 4.6 micron from 2014 through 2020 (Figure \ref{fig:WISE}), with data from June 2014 ($\sim$2 months after the April 2014 NOEMA observations) and June 2017 (taken within a week of the June 2017 NOEMA observations). The WISE photometry displays no significant variability, indicating that the disk irradiation may have remained relatively constant during that time, and therefore, that the millimeter variability is not tied to disk temperature changes. We can also rule out a change in disk temperature being responsible for the millimeter variability given that the $\sim$doubling of the millimeter flux in V1735 Cyg would imply  an equivalent $\sim$doubling in disk irradiation, which is unlikely.

\subsection{Gas and Dust Buildup in the Disk} \label{sec:buildup}

Because we see solely millimeter variability and no optical changes, one may speculate that this may be evidence that material is building up in the disk from the envelope. Using the equation

\begin{equation} \label{equ:Mcont}
    M_{cont} = \frac{gS_{\nu}d^2}{\kappa_{\nu}B_{\nu}(T)} \\
\end{equation}

 \noindent (which assumes an optically thin disk), where M$_{cont}$ is the continuum mass, $g=100$ is the gas-to-dust ratio, $S_\nu$ is the measured flux density at 2.7 mm, $d$ is the distance, $\kappa_{\nu}=0.2$ cm$^2$ g$^{-1}$ is the dust opacity coefficient at 2.7 mm, and $B_{\nu}(T)$ is the Planck function for a blackbody with a temperature of $T=30$ K, we find that the disk mass of V1735 Cyg \citep[at a distance $d=616$ pc;][]{Bailer2018} must have increased from 0.13 to 0.21 M$_{\odot}$ from 2014 to 2017. We note that this is consistent with a previous disk mass (0.20 M$_{\odot}$) estimated with SED modeling \citep{Gramajo2014}. Our measured disk mass change would correspond to a mass infall rate of 0.027 M$_{\odot}$ yr$^{-1}$ from the envelope, which is highly unlikely \citep{Ohtani2013, White2019}. Therefore, given the degree to which the continuum flux density changes, an unrealistic rate of mass infall would be necessary to account for the magnitude of the millimeter variability seen in V1735 Cyg. In addition, the (likely optically thin) C$^{18}$O emission was relatively constant from 2014 to 2017, implying that no C$^{18}$O has built up during that time. Thus, material buildup does not seem to be the source of the variability of V1735 Cyg at 2.7 mm.

\subsection{Free-Free Emission} 

One other potential source of millimeter variability is changes in the free-free emission of the system. Free-free emission can be identified by analysis of the spectral index, $\alpha$, of the millimeter emission. The more significant the free-free emission, the shallower the spectral index, down to -0.1--0.6 for purely free-free \citep{Reynolds1986}. Scattering in an optically thick disk can act to lower the spectral index as well, though this effect is generally strongest at the innermost regions of the disk \citep{Zhu2019, Liu2019}.

The change in $\alpha$ of V1735 Cyg can be measured using existing data from June 2013, April 2014, June 2018, and August 2018. \citet{Liu2018} weakly detected V1735 Cyg at 1.3 mm in June 2013. Given their 3-$\sigma$ upper limits, and using flux density estimates from April 2014 (F17), \citet{Liu2018} determined an upper limit on $\alpha$ of 1.7--2.0. Using the upper (106.7 GHz) and lower (91.3 GHz) sidebands described in Section \ref{sec:obs}, we are able to determine spectral indices of our June and August 2018 observations. We find tentative evidence of shallower slopes than \citet{Liu2018}, $\alpha=1.4\pm0.4$ in June 2018 and $\alpha=1.3\pm0.7$ in August 2018. These slopes are somewhat lower than the expected spectral index of most circumstellar disks, where generally $\alpha=2$--3 \citep{Beckwith1991, Ubach2012, Liu2018}, and are consistent with free-free emission. We note that the spectral indices we measure with our NOEMA data in V2494 Cyg ($\alpha=2.5$--2.6) and V2495 Cyg ($\alpha=2.3$--2.5) are in line with those of most circumstellar disks, thus free-free emission was likely not a significant contributor during those observations.

These possibly shallower spectral indices we find are suggestive that the slope of the millimeter emission of V1735 Cyg decreased from 2014 to 2017 while we see an increase in millimeter emission, and may indicate that the free-free emission of V1735 Cyg has increased to become a significant contributor to the overall SED near 2.7 mm. Free-free emission has been tied to ionized jets/winds in objects with disks \citep[e.g.,][]{Macias2016, Ubach2017, Espaillat2019}, and these jets/winds are linked to accretion \citep{Frank2014}. One would then expect to see signatures of accretion variability in V1735 Cyg which may be traced in the IR. However, the WISE photometry shows no significant variability between 2014 and 2017 (Figure \ref{fig:WISE}). It may be the case that the IR emission is variable due to accretion, but was not detected with the cadence of WISE.

\citet{Liu2018} note that free-free emission is not thought to be significant in FUor objects based on previous observations \citep[see][]{Rodriguez1990, Liu2014, Dzib2015, Liu2017}. This may indeed be the case for certain objects and/or during quiescent states without enhanced accretion, but free-free emission may become significant following an accretion event. As such, future observations of FUor objects would benefit not only from multi-epoch observations, but also from multiwavelength millimeter and centimeter observations \citep{Liu2017}. This will help inform how significant, if at all, free-free emission is for a given object. If significant, free-free emission may lead to overestimated disk masses.

\section{Summary} \label{sec:summary}

We observed six FUor objects (V1735 Cyg, V2494 Cyg, V2495 Cyg, V1057 Cyg, V1515 Cyg, and V733 Cep) in 2017 at 2.7 mm. Motivated by comparison to previously published works, we then followed up with coordinated 2.7 mm and optical (V, R, I) observations for three objects (V1735 Cyg, V2494 Cyg, and V2495 Cyg) to probe for flux variability. We did not see variability outside our stated uncertainties ($\sim$10--15\%) from 2017 to 2018 in either our millimeter or optical observations. However, we do see a $\sim80$\% increase in the 2.7 mm flux density of V1735 Cyg in our June 2017 data relative to archival April 2014 data from F17. Although we took steps to mitigate the effect of differing \textit{uv} coverages for each observation, it should be noted that they may still have had effects on our measurements.

We can likely rule out thermal changes in the disk as the source of millimeter variability in V1735 Cyg since 3.4 and 4.6 $\mu$m WISE photometry from 2014 to 2017 displayed no signs of corresponding variability, indicating that the millimeter variability is not related to temperature changes in the inner disk. Gas and dust buildup in the disk is also unlikely to be the sole mechanism behind the observed millimeter variability given that the mass transfer rate from the envelope to the disk necessary to account for the continuum flux density changes we see would be unreasonably large ($\sim0.027$ M$_{\odot}$ yr$^{-1}$). 

We find that the spectral slope of V1735 Cyg is shallower than expected for pure thermal dust emission at 3 mm, which may indicate a significant contribution from free-free emission. We also find that the 3 mm spectral index may have decreased since 2014, indicating a significant increase in the free-free emission. We hypothesize that V1735 Cyg may have experienced a small accretion event, leading to the ionization of ejected material, increasing the free-free emission and leading to the observed millimeter variability. If confirmed, this could imply that previously reported disk masses of FUor objects measured during enhanced accretion activity may be overestimated. Future study of FUor objects will benefit from both multi-epoch and multiwavelength observations to disentangle the free-free component from that of thermal dust emission and allow for more accurate disk mass estimates, which will help constrain what role thermal/gravitational instabilities have in triggering FUor outbursts. 

\acknowledgements

We thank the anonymous referee for a careful review and suggestions that greatly improved this paper. JW, CCE, and EM acknowledge support from the National Science Foundation under CAREER grant AST-1455042. \'AK acknowledges funding from the European Research Council under the European Union's Horizon 2020 research and innovation program under grant agreement 716155 (SACCRED). This work is based on observations carried out under project number SX17AG and S18AX with the IRAM NOEMA interferometer. IRAM is supported by INSU/CNRS (France), MPG (Germany), and IGN (Spain). These results made use of the Lowell Discovery Telescope (formerly Discovery Channel Telescope) at Lowell Observatory. Lowell is a private, nonprofit institution dedicated to astrophysical research and public appreciation of astronomy and operates the LDT in partnership with Boston University, the University of Maryland, the University of Toledo, Northern Arizona University, and Yale University. The Large Monolithic Imager was built by Lowell Observatory using funds provided by the National Science Foundation (AST-1005313). This publication makes use of data products from the Near-Earth Object Wide-field Infrared Survey Explorer (NEOWISE), a project of the Jet Propulsion Laboratory/California Institute of Technology. NEOWISE is funded by the National Aeronautics and Space Administration. This research made use of APLpy, an open-source plotting package for Python \citep{Aplpy}.

\vspace{5mm}
\facilities{IRAM:NOEMA, LDT}

\software{CASA, Python, Astropy, APLpy,}




\begin{thebibliography}{}

\bibitem[Ambartsumyan(1971)]{Ambart1971} Ambartsumyan, V.~A.\ 1971, Astrophysics, 7, 331

\bibitem[Armitage et al.(2001)]{Armitage2001} Armitage, P.~J., Livio, M., \& Pringle, J.~E.\ 2001, \mnras, 324, 705

\bibitem[Astropy Collaboration et al.(2018)]{Astropy2018} Astropy Collaboration, Price-Whelan, A.~M., Sip{\H{o}}cz, B.~M., et al.\ 2018, \aj, 156, 123

\bibitem[Audard et al.(2014)]{Audard2014} Audard, M., {\'A}brah{\'a}m, P., Dunham, M.~M., et al.\ 2014, Protostars and Planets VI, 387

\bibitem[Bailer-Jones et al.(2018)]{Bailer2018} Bailer-Jones, C.~A.~L., Rybizki, J., Fouesneau, M., et al.\ 2018, \aj, 156, 58

\bibitem[Beckwith \& Sargent(1991)]{Beckwith1991} Beckwith, S.~V.~W., \& Sargent, A.~I.\ 1991, \apj, 381, 250

\bibitem[Bell \& Lin(1994)]{Bell1994} Bell, K.~R., \& Lin, D.~N.~C.\ 1994, \apj, 427, 987

\bibitem[Bida et al.(2014)]{Bida2014} Bida, T.~A., Dunham, E.~W., Massey, P., et al.\ 2014, \procspie, 91472N

\bibitem[Bonnell \& Bastien(1992)]{B&B1992} Bonnell, I., \& Bastien, P.\ 1992, \apjl, 401, L31

\bibitem[Cieza et al.(2018)]{Cieza2018} Cieza, L.~A., Ru{\'\i}z-Rodr{\'\i}guez, D., Perez, S., et al.\ 2018, \mnras, 474, 4347

\bibitem[Clarke et al.(2005)]{Clarke2005}Clarke, C., Lodato, G., Melnikov, S.~Y., et al.\ 2005, \mnras, 361, 942


\bibitem[Cutri et al.(2012)]{Cutri2012} Cutri, R.~M. et al.\ 2012, VizieR Online Data Catalog, II/311

\bibitem[Dunham et al.(2012)]{Dunham2012} Dunham, M.~M., Arce, H.~G., Bourke, T.~L., et al.\ 2012, \apj, 755, 157

\bibitem[Dzib et al.(2015)]{Dzib2015} Dzib, S.~A., Loinard, L., Rodr{\'\i}guez, L.~F., et al.\ 2015, \apj, 801, 91


\bibitem[Espaillat et al.(2019)]{Espaillat2019} Espaillat, C.~C., Mac{\'\i}as, E., Hern{\'a}ndez, J., et al.\ 2019, \apjl, 877, L34


\bibitem[Feh{\'e}r et al.(2017)]{Feher2017} Feh{\'e}r, O., K{\'o}sp{\'a}l, {\'A}., {\'A}brah{\'a}m, P., et al.\ 2017, \aap, 607, A39

\bibitem[Frank et al.(2014)]{Frank2014} Frank, A., Ray, T.~P., Cabrit, S., et al.\ 2014, Protostars and Planets VI, 451

\bibitem[Gaia Collaboration (2018)]{Gaia} Gaia Collaboration\ 2018, VizieR Online Data Catalog, I/345

\bibitem[Gildas Team (2013)]{Gildas2013} Gildas Team\ 2013, GILDAS: Grenoble Image and Line Data Analysis Software, ascl:1305.010

\bibitem[Gramajo et al.(2014)]{Gramajo2014} Gramajo, L.~V., Rod{\'o}n, J.~A., \& G{\'o}mez, M.\ 2014, \aj, 147, 140

\bibitem[Hartmann \& Kenyon(1985)]{Hartmann1985} Hartmann, L., \& Kenyon, S.~J.\ 1985, \apj, 299, 462

\bibitem[Hartmann \& Kenyon(1996)]{Hartmann1996} Hartmann, L., \& Kenyon, S.~J.\ 1996, \araa, 34, 207

\bibitem[Harvey et al.(2008)]{Harvey2008} Harvey, P.~M., Huard, T.~L., J{\o}rgensen, J.~K., et al.\ 2008, \apj, 680, 495

\bibitem[Herbig(1977)]{Herbig1977} Herbig, G.~H.\ 1977, \apj, 217, 693

\bibitem[Johnson et al.(1966)]{Johnson1966} Johnson, H.~L., Mitchell, R.~I., Iriarte, B., et al.\ 1966, Communications of the Lunar and Planetary Laboratory, 4, 99


\bibitem[K{\'o}sp{\'a}l et al.(2016)]{Kospal2016} K{\'o}sp{\'a}l, {\'A}., {\'A}brah{\'a}m, P., Acosta-Pulido, J.~A., et al.\ 2016, \aap, 596, A52

\bibitem[Landolt(2009)]{Landolt2009} Landolt, A.~U.\ 2009, \aj, 137, 4186

\bibitem[Liu et al.(2014)]{Liu2014} Liu, H.~B., Galv{\'a}n-Madrid, R., Forbrich, J., et al.\ 2014, \apj, 780, 155

\bibitem[Liu et al.(2016)]{Liu2016} Liu, H.~B., Galv{\'a}n-Madrid, R., Vorobyov, E.~I., et al.\ 2016, \apjl, 816, L29

\bibitem[Liu(2019)]{Liu2019} Liu, H.~B.\ 2019, \apjl, 877, L22

\bibitem[Liu et al.(2018)]{Liu2018} Liu, H.~B., Dunham, M.~M., Pascucci, I., et al.\ 2018, \aap, 612, A54.

\bibitem[Liu et al.(2017)]{Liu2017} Liu, H.~B., Vorobyov, E.~I., Dong, R., et al.\ 2017, \aap, 602, A19

\bibitem[Mac{\'\i}as et al.(2016)]{Macias2016} Mac{\'\i}as, E., Anglada, G., Osorio, M., et al.\ 2016, \apj, 829, 1

\bibitem[Magakian et al.(2013)]{Magakian2013} Magakian, T.~Y., Nikogossian, E.~H., Movsessian, T., et al.\ 2013, \mnras, 432, 2685

\bibitem[McMullin et al.(2007)]{McMullin2007} McMullin, J. P., Waters, B., Schiebel, D., et al.\ 2007, Astronomical Data Analysis Software and Systems XVI (ASP Conf. Ser. 376), ed. R. A. Shaw, F. Hill, \& D. J. Bell (San Francisco, CA: ASP), 127


\bibitem[Ohtani \& Tsuribe(2013)]{Ohtani2013} Ohtani, T., \& Tsuribe, T.\ 2013, \pasj, 65, 93

\bibitem[Peneva et al.(2009)]{Peneva2009} Peneva, S.~P., Semkov, E.~H., \& Stavrev, K.~Y.\ 2009, \apss, 323, 329

\bibitem[Pety(2005)]{Pety2005} Pety, J.\ 2005, SF2A-2005: Semaine De L'astrophysique Francaise, 721




\bibitem[Reipurth et al.(2007)]{PP6} Reipurth, B., Jewitt, D., \& Keil, K., 2007, Protostars and Planets V

\bibitem[Reynolds(1986)]{Reynolds1986} Reynolds, S.~P.\ 1986, \apj, 304, 713

\bibitem[Robitaille \& Bressert(2012)]{Aplpy} Robitaille, T., \& Bressert, E.\ 2012, APLpy: Astronomical Plotting Library in Python, ascl:1208.017

\bibitem[Rodriguez et al.(1990)]{Rodriguez1990} Rodriguez, L.~F., Hartmann, L.~W., \& Chavira, E.\ 1990, \pasp, 102, 1413


\bibitem[Siwak et al.(2018)]{Siwak2018} Siwak, M., Winiarski, M., Og{\l}oza, W., et al.\ 2018, \aap, 618, A79

\bibitem[Takagi et al.(2018)]{Takagi2018} Takagi, Y., Honda, S., Arai, A., et al.\ 2018, \aj, 155, 101

\bibitem[Takami et al.(2019)]{Takami2019} Takami, M., Chen, T.-S., Liu, H.~B., et al.\ 2019, \apj, 884, 146

\bibitem[Ubach et al.(2012)]{Ubach2012} Ubach, C., Maddison, S.~T., Wright, C.~M., et al.\ 2012, \mnras, 425, 3137

\bibitem[Ubach et al.(2017)]{Ubach2017} Ubach, C., Maddison, S.~T., Wright, C.~M., et al.\ 2017, \mnras, 466, 4083

\bibitem[Vorobyov \& Basu(2005)]{Vorobyov2005} Vorobyov, E.~I., \& Basu, S.\ 2005, \apjl, 633, L137


\bibitem[White et al.(2019)]{White2019} White, J.~A., K{\'o}sp{\'a}l, {\'A}., Rab, C., et al.\ 2019, \apj, 877, 21

\bibitem[Zhu et al.(2010)]{Zhu2010} Zhu, Z., Hartmann, L., Gammie, C.~F., et al.\ 2010, \apj, 713, 1134

\bibitem[Zhu et al.(2019)]{Zhu2019} Zhu, Z., Zhang, S., Jiang, Y.-F., et al.\ 2019, \apjl, 877, L18

\end{thebibliography}
\end{document}